# Accelerometer-Based Multivariate Time-Series Dataset for Calf Behavior Classification

**Authors**

Oshana Iddi Dissanayake*[1,3], Sarah E. McPherson[3,4,5], Joseph Allyndrée[2,3], Emer Kennedy[3,4], Pádraig Cunningham[1], Lucile Riaboff[1,3,6]

**Affiliations**

1. School of Computer Science, University College Dublin, Ireland.
2. School of Maths and Stats, University College Dublin, Ireland.
3. VistaMilk SFI Research Centre, Ireland.
4. Teagasc, Animal & Grassland Research and Innovation Centre, Moorepark, Fermoy, Co. Cork, P61C997, Ireland.
5. Animal Production Systems Group, Wageningen University & Research, Wageningen, The Netherlands.
6. GenPhySE, Université de Toulouse, INRAE, ENVT, 31326, Castanet-Tolosan, France.

**Corresponding author's email address and Twitter handle**

oshana.dissanayake@ucdconnect.ie / @OshanaDiss

**Abstract**

Getting new insights on pre-weaned calf behavioral adaptation to routine challenges (transport, group relocation, etc.) and diseases (respiratory diseases, diarrhea, etc.) is a promising way to improve calf welfare in dairy farms. A classic approach to automatically monitoring behavior is to equip animals with accelerometers attached to neck collars and to develop machine learning models from accelerometer time-series. However, for this accelerometer time-series data to be used for model development, it must be equipped with labels describing behaviors (gold standard). Obtaining these labels requires annotating behaviors from direct observation or videos, a time-consuming and labor-intensive process. In addition, accurate alignment between accelerometer data and behaviors is always challenging due to synchronization issues. To address this challenge, we propose the **ActBeCalf** (*Accelerometer Time-Series for Calf Behaviour classification*) dataset: 30 pre-weaned dairy calves (Holstein Friesian and Jersey) housed in 4 group pens at Teagasc Moorepark Research Farm (Fermoy, Co. Cork, Ireland) were equipped with a 3D-accelerometer sensor (sampling rate: 25 Hz) attached to a neck-collar from one week of birth for 13 weeks. The calves were simultaneously filmed with a high-up camera in each pen. Every 15 days, accelerometers were removed from the collars to recharge the battery and reattached to the neck collar again. At the end of the trial, behaviors were manually annotated from the videos using the Behavioral Observation Research Interactive Software (BORIS) by 3 observers (Cohen's Kappa = 0.72 ± 0.01) using an ethogram with 23 pre-weaned dairy calves' behaviors. Observations were synchronized with the accelerometer timestamps using an external clock and aligned to the corresponding accelerometer time-series. The synchronization has also been manually inspected for each time-series. Thereby, ActBeCalf contains 27.4 hours of accelerometer data from 30 calves (age 23.7 ± 10.7 days) aligned adequately with calf behaviors. The dataset includes the main behaviors, like lying, standing, walking, and running, as well as less prominent behaviors, such as sniffing, scratching, social interaction, and grooming. Finally, ActBeCalf was used for behavior classification with machine learning models to demonstrate its reliability. For that purpose, we developed two machine learning models using (i) two classes of behaviors, [active and inactive; model 1] and (ii) four classes of behaviors [running, lying, drinking milk, and every other behavior grouped into the "other" class; model 2]. We got a balanced accuracy of 92% [model 1] and 84% [model 2]. The code utilized for the classification is publicly available in the dataset repository. Therefore, ActBeCalf is a comprehensive and ready-to-use dataset for classifying pre-weaned calf behavior from the accelerometer time-series.

## SPECIFICATIONS TABLE

| Subject | Applied Machine Learning |
|---|---|
| Specific subject area | Calf behavior classification using accelerometer time-series data. |
| Type of data | Table with raw accelerometer data and annotated labels. |
| Data collection | This experiment involved 30 Holstein Friesian and Jersey pre-weaned calves housed in 4 group pens. Each calf was equipped with a 3D accelerometer sensor (AX3, Axivity Ltd, Newcastle, UK; 11g) sampled at 25 Hz and attached to a neck collar from 1 week of birth over 13 weeks. The sensors were removed bi-weekly to recharge the batteries and extract the data. An 8Mp Dome3 CCTV camera was fixed high up in each pen. Behaviors were manually annotated from the videos using the Behavioral Observation Research Interactive Software based on an ethogram (24 behaviors). After timestamp synchronization, observations and accelerometer time-series were aligned using Python (v3.9). No normalization has been applied to the accelerometer time-series. |
| Data source location | Data were collected at Teagasc Moorepark Research Farm, Fermoy, Co. Cork, Ireland (50°07′N; 8°16′W). |
| Data accessibility | Repository name: **Accelerometer-Based Multivariate Time-Series Dataset for Calf Behavior Classification.**<br>Data identification number: **10.5281/zenodo.13259482**<br>Direct URL to data: **https://zenodo.org/records/13259482** |
| Related research article | |

## VALUE OF THE DATA

- **Support research in pre-weaned calf behaviour classification from accelerometer data**:
  Compared with cows, pre-weaned calf behavior classification from accelerometer data has been little studied. However, developing models for automatically classifying calf behavior from accelerometer data would be beneficial for measuring behavioral adaptation to the routine challenges experienced by pre-weaned dairy calves (transport, pen relocation, dehorning, weaning, respiratory diseases, diarrhea, etc.). However, one major challenge is the need for datasets with accelerometer time-series aligned with the behaviors carried out by the animals, manually annotated from videos or direct observations. Indeed, this step is highly time-consuming and labor-intensive. Time-synchronization between accelerometer data and annotated behaviors is also challenging due to the different real-time clocks between accelerometer sensors and cameras. ActBeCalf contains 27.4 hours of annotations with 24 behaviors carefully synchronized with accelerometer data collected from 30 calves. ActBeCalf is thus a comprehensive and ready-to-use dataset that will speed up the research in classifying calf behavior from accelerometer data.

- **Address the methodological challenges in livestock ruminant behaviour classification from accelerometer data:**
  ActBeCalf is a suitable dataset for developing and validating machine learning models for the classification of livestock ruminant behavior while addressing the current limitations identified in the field, namely (1) the lack of generalizability of the models when applied to new animals and (2) the difficulty of classifying a broad spectrum of behaviors, including those that are only occasionally observed [6][10]. ActBeCalf encompasses 30 animals with 27.4 hours of observation, focusing on 24 pre-weaned calf behaviors. This comprehensive dataset allows for two key purposes: (1) training models on a subset of the animals while testing on others to assess the generalizability of the models, and (2) evaluating the models in challenging scenarios by including rarely observed behaviors in the classification process. This helps identify the optimal configuration for time-series normalization and filtering, feature extraction, data augmentation, time-series segmentation, and machine-learning algorithm modeling.

- **Standardize accelerometer configuration for collecting accelerometer data from neck-collars in cattle:**
  Data are usually collected using protocols specific to the study (e.g., the position of the accelerometer on the animal body, the direction of the accelerometer axes, sampling rate, etc.). The lack of reproducibility from one study to another is a limitation to collaborative research, as it does not allow datasets to be pooled for larger volumes without additional manual annotations, nor does it allow signal processing and modeling techniques to be compared objectively to make recommendations for future studies. Furthermore, the time-synchronization between accelerometer data and annotated behaviors is highly challenging, but the literature needs to describe better how to solve this issue. In this respect, ActBeCalf can help researchers standardize their experimental design so that it can be used to enhance their dataset. Thereby, ActBeCalf supports collaborative efforts among researchers in the field, thus accelerating advancements in cattle behavior classification from accelerometer data.

- **Support development in Multivariate Time-Series Classification:**
  ActBeCalf is a ready-to-use dataset for classifying a label (gold standard) from 3 raw accelerometer time-series (X, Y, Z). Thus, ActBeCalf supports fundamental development in Multivariate Time-Series Classification. ActBeCalf should also promote interdisciplinary collaborations between animal scientists, ethologists, and computer scientists to break down barriers in animal behavior classification from sensor data while developing new techniques in applied Machine Learning.

# BACKGROUND

Assessing the impact of routine challenges (e.g., separation from dams, transport, dehorning) on the welfare of pre-weaned calves is a promising way to make recommendations for reducing calf distress in dairy farms. In that regard, monitoring changes in behavior in pre-weaned calves is relevant as the number and duration of behavior bouts may be altered after a stressful event [11]. Behavior can be monitored automatically from accelerometer data, but developing a machine-learning model is necessary to classify behaviors accurately [6], [10]. ActBeCalf has been produced in that context, combining accelerometer data aligned with annotated behaviors. It contains prominent and non-prominent behaviors, allowing one to focus on the main behaviors of the time budget (e.g., lying, drinking milk) or certain key behaviors (e.g., running). Furthermore, ActBeCalf encompasses 30 calves so that the model can be trained and validated adequately to assess its generalisability from one animal to another, a methodological challenge that is highlighted in the literature [5], [10]. Our data article adds value to our original research article by supporting model development for applicative research in livestock ruminant behavior and fundamental research in multivariate time-series classification.

# DATA DESCRIPTION

The dataset consists of a single **CSV** file containing 7 columns. The columns are as follows:

**Table 01:** AcTBeCalf dataset column names and descriptions.

| Column name in the AcTBeCalf | Description |
| --- | --- |
| dateTime | Timestamp of the accelerometer reading, with the accelerometer sampling rate set to 25 Hz. |
| calfid | Identification number of the calf from which the data was collected (1-30). |
| accX | The accelerometer reading for the axis X (top-bottom direction; see Figure 03). |
| accY | The accelerometer reading for the axis Y (backward - forward direction; see Figure 03). |
| accZ | The accelerometer reading for the axis Z (left-right direction; see Figure 03) |
| behavior | Annotated behavior for each reading based on the ethogram of 23 behaviors (see Table 02). |
| segId | Segment identification number associated with each accelerometer reading/row. A segment is an accelerometer time-series associated with a single behavior annotated on the same calf (see Figure 01). Rows with the same segId represent all the accelerometer readings of the same segment. |

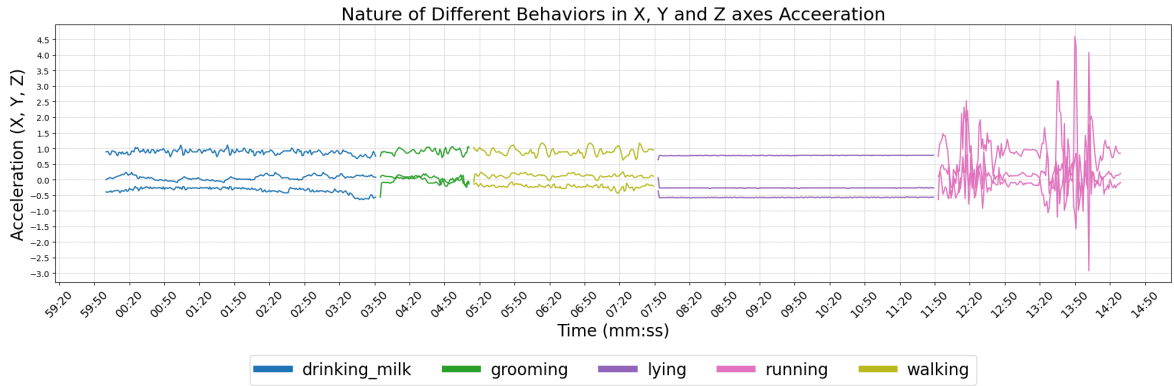

**Figure 01:** Behavior segments in a continuous annotation sequence. Different segments are shown in different colors.

Table 02 presents details about the ethogram used for annotation purposes. Some behaviors have a modifier that clarifies the state or characterizes the behavior. *Out-of-frame* annotation is used only for noting the animal's state and is not considered a behavior.

**Table 02:** Ethogram with the 23 behaviours used for the annotation process.

| Behaviour | Modifiers | Description |
| --- | --- | --- |
| Standing | | Animal is in a static upright standing position with weight placed on all four legs. |
| Lying | | Animal rests sternally or laterally, with all four legs hunched close to the body, and is either awake or asleep. |
| Defecation | standing, lying | Animal is defecating. |
| Urination | standing, lying | Animal is urinating. |
| Rumination | standing, lying | Animal is ruminating. |
| Drinking | milk, water, electrolytes, fail | Animal is drinking: milk, water, electrolytes. |
| Rising | | Animal is in the process of rising from a lying-down position. |
| Lying down | | Animal is in the process of lying-down from a standing position. |
| Grooming | standing, lying | Animal uses its tongue to repeatedly lick its back, side, leg, tail areas. |

| Behavior | Modifiers | Description |
|---|---|---|
| Social interaction | groom, nudge, sniff \| posture: standing, lying | Animal engages in social interaction with another animal. |
| Play | headbutt, object, jump, mount \| posture: standing, lying | Animal runs, jumps, changes direction suddenly, bucks, kicks hind legs, twists or rotates body / Animal mounts, or attempts to mount, a pen mate / Animal is engaged in head-to-head pushing with another animal / Calf plays with an object in the pen. |
| HOP | standing, lying | Calf puts the tip of its nose/head or more out of the pen through an opening. |
| Oral manipulation of pen | standing, lying | Animal licks, nibbles, sucks, or bites at the pen structure (barriers, walls, buckets, troughs etc.). |
| Sniff | standing, lying, walking | Animal sniff the ground or any part of the pen structure. |
| Abnormal | tongue rolling, urine drinking, cross-suckle naval, cross-suckle udder, cross-suckle other \| posture: standing, lying | Animal performs an abnormal behavior: tongue rolling (Animal makes repeated movements with its tongue inside or outside its mouth); urine drinking (drinks urine of another calf); naval (animal sucks on navel of another animal); cross-suckling (animal attempts to suck the udder or any other part of another animal). |
| SRS | scratch, rub, stretch \| posture: standing, lying | Animal scratches itself with one of their legs (generally hind legs); Animal rubs itself on pen structure; animal stretches itself. |
| Eating | forage, concentrates, bedding, other \| posture: standing, lying | Animal is eating: forage, concentrates, bedding, other. |
| Vocalization | standing, lying | Animal is visibly vocalizing. |
| Walking | forward, backward | Animal is walking or shuffling about: forward, backward. |
| Run | play, not play | Animal is running. |
| Pacing | | Animal is pacing. |
| Cough | standing, lying | Animal is visibly coughing. |
| Fall | | Calf falls down due to tripping, slipping, etc. |
| Out-of-frame | standing, lying | Animal is out of frame and cannot tell what they are doing. |

Figure 02 illustrates the proportions of individual behaviors in the dataset and the number of calves from which the behavior data was collected. The dataset includes 21 out of the 23 behaviors listed in the ethogram; HOP and Pacing behaviors were not observed in the analyzed videos.

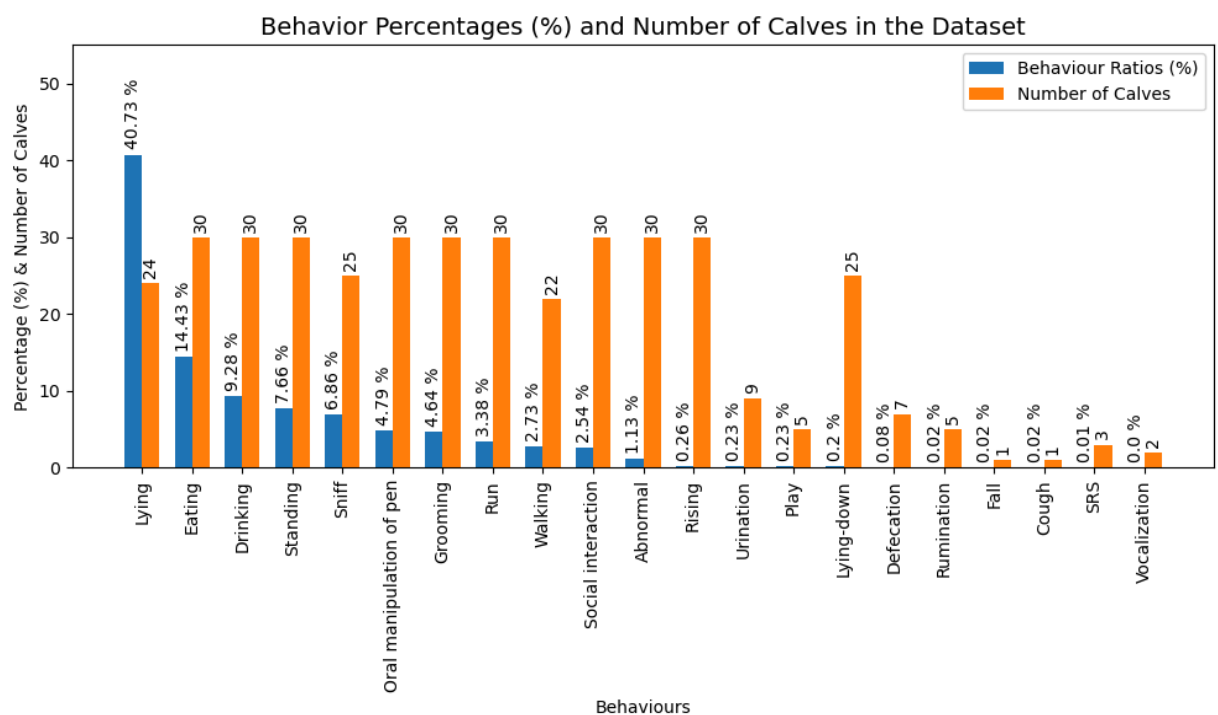

**Figure 02:** Behavior duration as a percentage (%) of the total dataset.

# EXPERIMENTAL DESIGN, MATERIALS AND METHODS

## 1. Location

The experiment was conducted at the Teagasc Moorepark Research Farm, Fermoy, Co. Cork, Ireland (50°07′N; 8°16 134′W) from January 21st to April 5th, 2022. The trial was carried out per the European Union (Protection of Animals Used for Scientific Purpose) Regulations 2012 (S.I. No. 543 of 2012), and the ethical approval was obtained from the Teagasc Animal Ethics Committee (TAEC; TAEC2021–319). 30 Irish Holstein Friesian and Jersey pre-weaned calves were utilized for the experiment. The calves were managed according to conventional rearing and management practices [2] at Teagasc Moorepark Research Farm.

## 2. Animal management

Within one hour after birth, the calves and dams were separated after calving. The calves were then moved to straw-bedded pens, and their mother's colostrum was artificially fed (within <2h post-birth) at a rate of 8.5% of their birth weight. Preceding this process, the calves were fed transition milk from their dams at a rate of 10% of their birth weight, administered twice daily for their subsequent five feedings. Following the transition milk, the calves were given 2.5 liters of milk replacer (26% crude protein; Volac Heiferlac Instant, Volac, Hertfordshire, UK) twice daily. From 3 to 7 days of age, the calves were moved to a group pen (see Figure 03) and fed with an automatic milk feeder at a rate of 6 liters per calf per day, with unlimited access to hay, concentrates, and water. The calves were gradually weaned at 56 days using the automatic feeder.

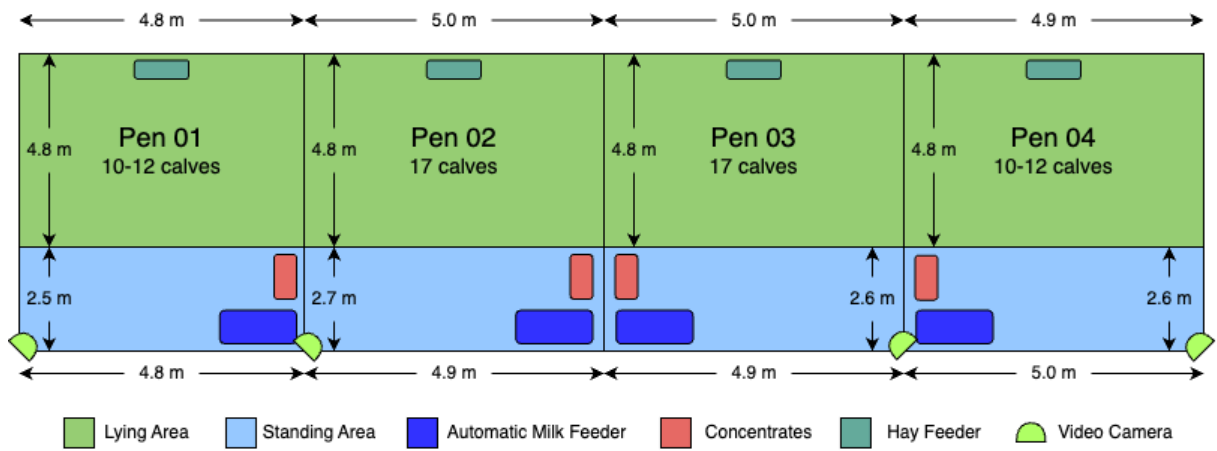

**Figure 03:** Grouped Pen layout.

## 3. Accelerometer data collection

Each calf was equipped with a 3D accelerometer data logger (Axivity LTD, Newcastle, UK, https://axivity.com/product/ax3) attached to a neck collar starting one week after birth until two months of age (see Figure 04). The accelerometers were configured using the OmGUI software (https://github.com/digitalinteraction/openmovement/wiki/AX3-GUI) with a sensitivity of ±8 g and a sampling rate of 25Hz, ensuring a battery life of ≈27 days (https://axivity.com/userguides/ax3/). Data was stored on a 512Mb NAND flash memory. The accelerometers measured 23×32.5×7.6 mm and weighed 11g (see Figure 05). Each sensor was wrapped in cling film and cotton wool, then secured to the collar with vet wrap and insulating tape, and positioned on the left side of the neck in the same orientation for all calves (see Figure 04). The X-axis detected the top-bottom direction, the Y-axis detected the backward-forward direction, and the Z-axis detected the left-right direction. Collars were tightly adjusted, with a 13g metal ring added to prevent movement from the designated side. Collars were removed every two weeks over the next 10 weeks to retrieve data format using OmGUI software. The accelerometer data was extracted in CWA format and stored on a hard drive before being backed up on a storage server. The accelerometers were then recharged and relaunched over a few hours. Following this, the collars were reassembled and reapplied to the calves. Each calf was fitted with the same accelerometer sensor identifier for 10 weeks.

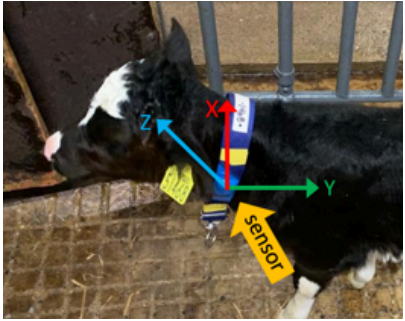
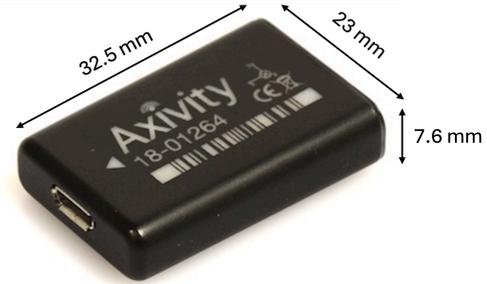

**Figure 04:** Accelerometer fixture and the sensor orientation.

**Figure 05:** AX3 Accelerometer.

## 4. Video collection

In addition to the accelerometer data collection, a set of DVRs (PRIMA XR5 8MP 4K; Equicom Limited, Cobh, Co. Cork, Ireland) and four Varifocal Dome CCTV (records up to 25 frames per second) Camera with 40 m night vision (Equicom Limited, Cobh, Co. Cork, Ireland) mounted in each pen (approximately 2.7 meters above the ground) were used to record videos of the calves (see Figure 03 and Figure 06). Videos were continuously recorded throughout the experiment as the DVR was replaced with a similar DVR, depending on its storage utilization. Videos were then extracted and converted into *AVI* format using VideoProc software (https://www.videoproc.com/). A total of 2092 hours of video footage was collected by the end of the trial.

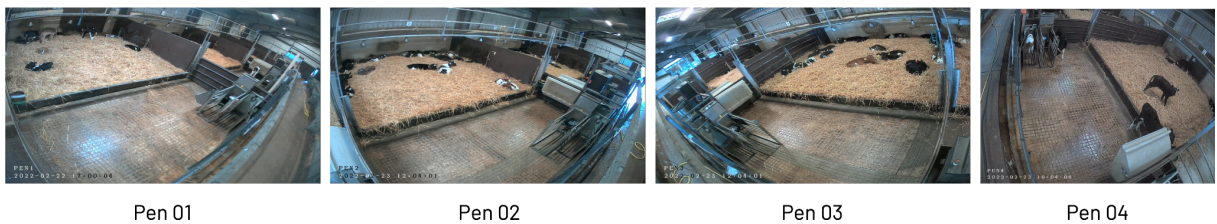

**Figure 06:** Video camera footage from pen 01, pen 02, pen 03 and pen 04.

## 5. Behaviour annotation

### 5.1. Video selection

The videos were recorded for each camera as 1-hour segments with the H.265 video compression format. A set of 3 ± 2 hours-videos has been selected for each calf to encompass a broad spectrum of behaviors for each animal. Since pre-weaned calves spend most of their time lying down, an algorithm was developed using accelerometer data to select at least one 1-hour video segment where the calf was active. The algorithm filtered the accelerometer data corresponding to each video and calculated the magnitude of acceleration (see Equation 01) for each timestamp. 1G is reduced to account for the gravitational acceleration [12]. Magnitude readings greater than 0.5 were labeled as active. If the proportion of active readings to total readings exceeded 80%, that video segment was classified as active (see Figure 07). The 80% ratio was chosen to identify videos consistently exhibiting more active behaviors throughout their duration. This algorithm ensures that at least one hour of the video is selected when the calf is mainly active so as not to annotate exclusively lying behavior.

$$Magnitude \ = \ max(\sqrt{(accX^2 \ + \ accY^2 \ + \ accZ^2}\ - \ 1g, \ 0) \hspace{2cm} \text{Equation 01}$$

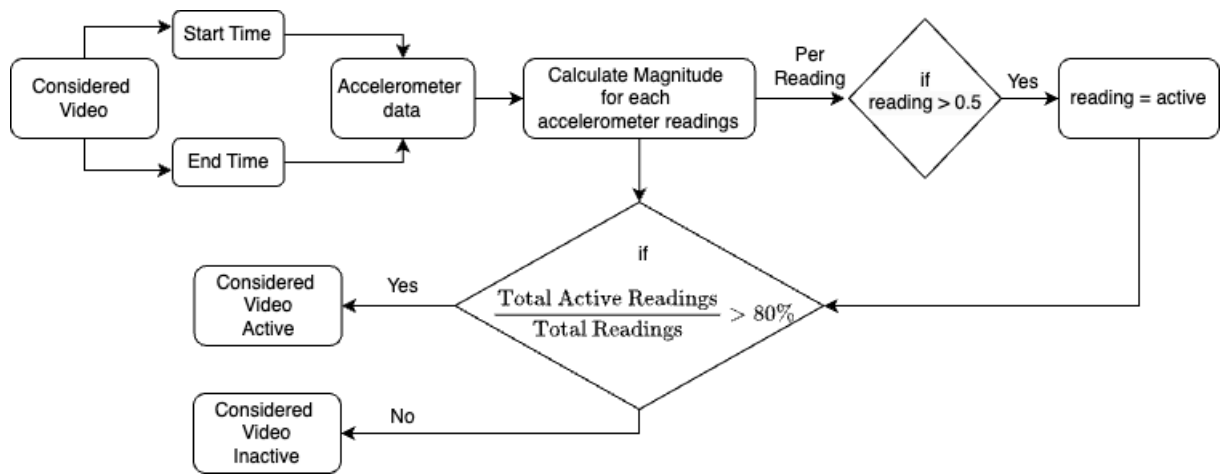

**Figure 07:** Active Inactive video separation algorithm.

## 5.2. Annotations with BORIS software

The original AVI videos were converted to MP4 format to ensure compatibility with the BORIS software. The conversion was done using a Python script, and it is available in the *holsteinlib* folder under the name *video_avi_to_mp4.py*. Each set of selected videos was annotated using the Behavioural Observation Research Interactive Software (BORIS) [8]. Three annotators conducted the annotations. Initial BORIS observations were conducted as training to assess the consistency and reliability of the annotators. These training observations were not included in the final analysis. During the training, annotators frequently consulted each other to ensure a shared understanding of the 23 behavior definitions. At the end of the training, the three annotators independently annotated a one-hour video continuously of a single calf. Concordance within the one-hour video was measured every second across the 23 behaviors, resulting in a Cohen's Kappa average of 0.72 ± 0.01, thereby validating the consistency between the 3 annotators for the rest of the process. The 30 calves and videos selected for annotation for each calf were then distributed among the three annotators. Each annotator then carried out his BORIS observation, ensuring that a single annotator annotated each calf for at least 15 minutes. In total, 27.4 hours of observations were completed for the 30 calves (0.91 +/- 0.46 hours of observation per calf). The age of calves during the annotation process was 23.7 + / − 10.7 days. The annotation process using BORIS was performed as follows:

(1) **Create a BORIS project**: An exhaustive ethogram comprising 23 behaviors, adapted from [1] for this experiment, was established. Additionally, modifiers were included to provide a more detailed description of certain behaviors, such as information on posture, feeding, and interaction. The ethogram and modifiers with the definition of each behavior are shown in Table 02. Each behavior has been set up as a "state" behavior in BORIS so that there is a start and stop time associated with each behavior. This enables the time-series to align with this start and end time after synchronizing the accelerometer timestamps with behaviors annotated from the videos (see section 6). Keyboard shortcuts associated with each behavior were defined. The unique identifier of each of the 30 calves was also input into the BORIS project.

(2) **Initiate a BORIS observation**: For each calf, each video selected was viewed separately, annotating the calf's behavior. A BORIS observation, therefore, corresponds to a single video for a single calf. The start date and time initialized for each BORIS observation were the start date and time of the video. As per the other configurations, the time formation was set to seconds. Each observation was given a unique ID composed of the calf ID, video ID, and annotator ID. A screenshot of the BORIS interface when initiating an observation is displayed in Figure 08.

(3) **Create a BORIS observation**: For each observation, the calf whose behavior was to be annotated was first identified based on photos of the calf, the collar color, and the tape attached to one of the hind legs (see Figures 09 and 10). Once the calf had been identified in the video, the behaviors were recorded using keyboard shortcuts associated with the ethogram. Behaviors were only recorded when the calf was easily observable, and the behavior could be clearly identified. The entire video was viewed in an attempt to capture rare or atypical behavior. A screenshot of a BORIS observation in progress is presented in Figure 11. BORIS observation was then closed. The BORIS software automatically notified any inconsistency in the times associated with each recorded behavior and manually inspected for correction.

(4) **Export a BORIS observation**: Each BORIS observation was exported using the "aggregated event" option so that a CSV file was produced for each observation (see Figure 12).

## 6. Merging annotations with accelerometer time-series

Accelerometer time-series were aligned with the annotated behaviors based on the timestamps of the accelerometer records and the time recorded in the BORIS observations. As the real-time clocks of the accelerometer sensors and cameras differ, the alignment was accomplished using a common external clock from a smartphone, which was called the *reference time*. All the codes used in that step are available in the *Data Preprocessing* folder.

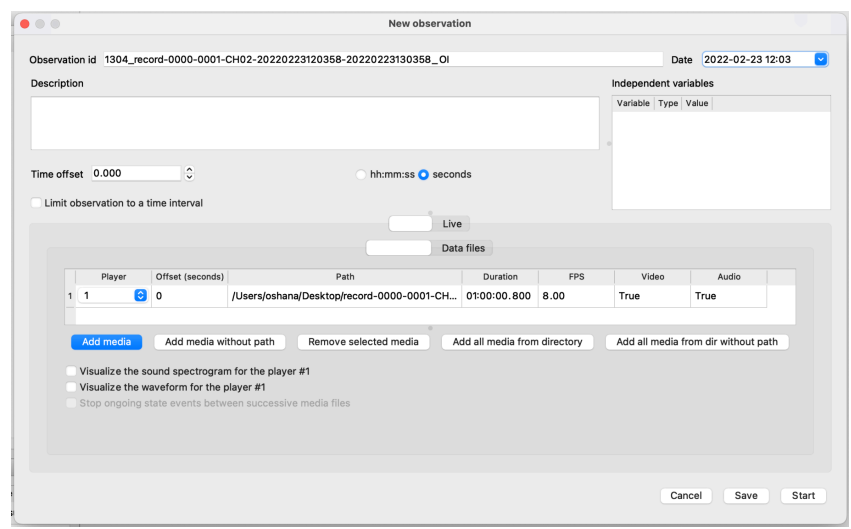

**Figure 08:** Importing a video into BORIS.

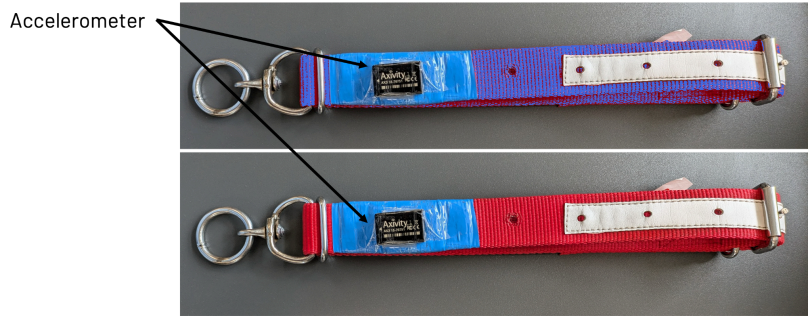

**Figure 09:** Accelerometer mounted on the neck-collar.

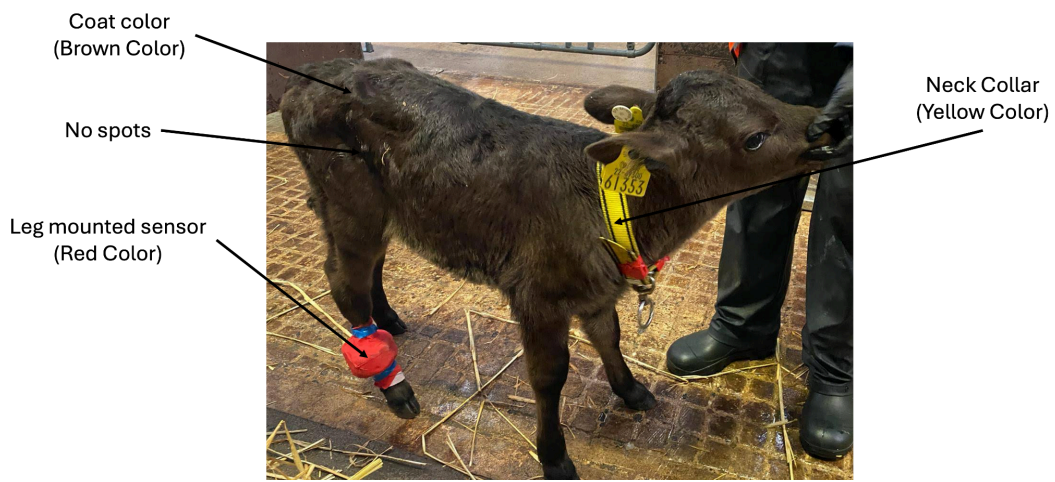

**Figure 10:** Calf identification characteristics

**Figure 11:** BORIS Interface (conducting annotation).

| Observation id | Observation date | Media file | Total length | FPS | Subject | Behavior | Behavioral category | Modifiers | Behavior type | Start (s) | Stop (s) | Duration (s) | Comment start | Comment stop |
|---|---|---|---|---|---|---|---|---|---|---|---|---|---|---|
| 1452_record-00 | 23/02/2022 12:04 | /Users/oshana/Desktop/ | 3600.85 | 12 | 1452 | drinking | Feeding | milk | STATE | 0 | 145.153 | 145.153 | | |
| 1452_record-00 | 23/02/2022 12:04 | /Users/oshana/Desktop/ | 3600.85 | 12 | 1452 | walking | Normal | Backward | STATE | 176.655 | 181.157 | 4.502 | | |
| 1452_record-00 | 23/02/2022 12:04 | /Users/oshana/Desktop/ | 3600.85 | 12 | 1452 | standing | Position | | STATE | 196.153 | 200.405 | 4.252 | | |
| 1452_record-00 | 23/02/2022 12:04 | /Users/oshana/Desktop/ | 3600.85 | 12 | 1452 | walking | Normal | Forward | STATE | 201.734 | 206.487 | 4.753 | | |
| 1452_record-00 | 23/02/2022 12:04 | /Users/oshana/Desktop/ | 3600.85 | 12 | 1452 | standing | Position | | STATE | 253.906 | 264.656 | 10.75 | | |
| 1452_record-00 | 23/02/2022 12:04 | /Users/oshana/Desktop/ | 3600.85 | 12 | 1452 | standing | Position | | STATE | 287.072 | 295.318 | 8.246 | | |
| 1452_record-00 | 23/02/2022 12:04 | /Users/oshana/Desktop/ | 3600.85 | 12 | 1452 | grooming | Normal | standing | STATE | 296.072 | 311.073 | 15.001 | | |

**Figure 12:** Format of an output CSV record from BORIS.

### 6.1. Aligning the accelerometer time-series to the reference time

In order to align the accelerometer time-series to the reference time, a pattern has been created in the accelerometer signal while the associated reference time was noted. The pattern must be sensitive to be systematically detected by an algorithm and specific to avoid confusion with patterns resulting from the acceleration of the animal's body. For that purpose, the following procedure has been applied during data collection: Every 15 days before attaching the collars, accelerometers were manually shaken 5 times for 5 seconds with 10 seconds rest between each shake. The reference time to the nearest second at the time of the first shake was noted. An algorithm was then developed with Python (v.3.9) to automatically detect the pattern generated by the procedure (see Figure 13, see the code *shake_pattern_detector.ipynb*). The performance of this algorithm was tested until we got 100% true positives and 0% false positives on 100 accelerometer time-series randomly selected. The gap in seconds between the accelerometer timestamp recorded at the first shake and the corresponding reference time was then calculated. The gap was corrected over the entire accelerometer time-series so that the accelerometer data aligned with the reference time (see the code *accelerometer_time_correction.ipynb*).

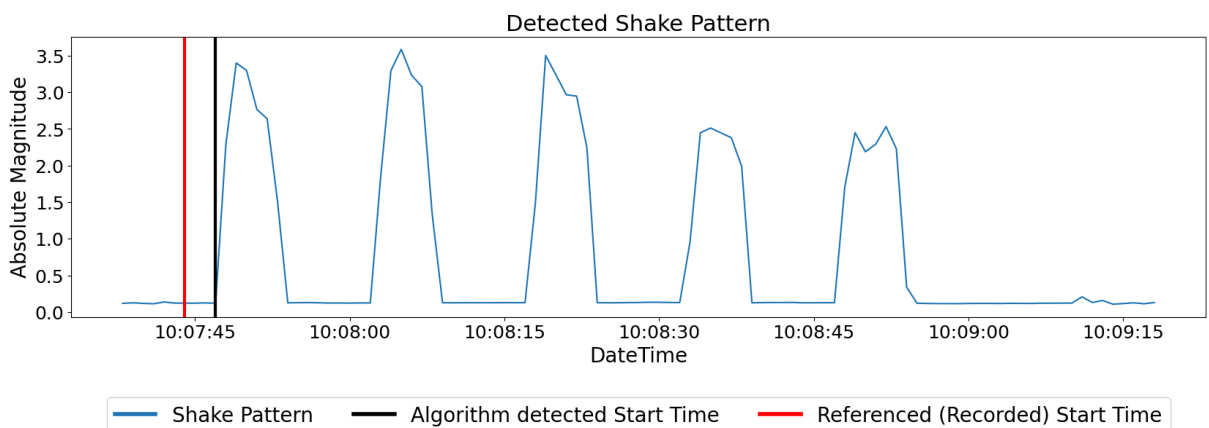

**Figure 13:** Automatically detected shake pattern by the algorithm in the smoothed accelerometer signal.

### 6.2. Correcting the accelerometer time drift

As explained in the documentation of the AX3 accelerometer sensors, the internal clock drifts slightly (estimation: 0.18 seconds per hour; https://axivity.com/userguides/ax3/settings/). An additional correction was thus applied to take the time drift into account. For that purpose, accelerometers were also shaken 5 times for 5 seconds with 10 seconds rest between each shake after removing the accelerometer sensors every 15 days. The reference time was recorded as explained in the section 6.1. The algorithm was then used to detect the pattern at the beginning and end of each accelerometer time-series (see Figure 13). The time drift of the accelerometer sensor was then computed based on (i) the actual duration recorded between the pattern at the beginning of the accelerometer time-series and the pattern at the end of the accelerometer time-series using the reference time and (ii) the detected duration recorded with the accelerometer timestamp. The time drift was then corrected for each accelerometer record, considering both the reference time alignment and the accelerometer time drift ensuring the alignment of the launch shake pattern reference time with the detected time as well as the alignment of the stop shake pattern reference time with the detected time at the same time (see the code accelerometer_time_correction.ipynb).

### 6.3. Aligning the annotations to the reference time

The smartphone with the reference time was periodically displayed to the camera to synchronize its timestamps. When a video is chosen for annotation, the time difference between the video and the reference time is calculated by examining the reference time recorded on the corresponding date and camera. This gap is then adjusted accordingly to the corresponding BORIS observation. No time drift has been observed in the camera's internal clock.

### 6.4 Aligning the accelerometer time-series with annotations

After aligning the accelerometer time-series time stamps with the reference time, correcting the accelerometer time drift, and aligning the video annotations to the reference time, the accelerometer time-series were aligned with the annotations, thus ensuring correct alignment (see the code *aligning_accelerometer_data_with_annotations.ipynb*).

### 6.4. Manual inspection of the accelerometer time-series aligned with annotated behaviours

Manual inspection of the time-series involved several key steps to ensure the alignment between the accelerometer data and annotated behaviors with accuracy at the one second level. The visual inspection of the accelerometer time-series with the annotated behaviors allows for confirming that the alignment is accurate and correcting it manually (+/- X seconds) if necessary. The manual inspection was done as follows (Refer to the code *manual_inspection_ts_validation.ipynb*.):

1. **Select the accelerometer time-Series associated with observation**: The accelerometer time-series corresponding to each Boris observation was identified and selected.
2. **Plot accelerometer time-Series with annotated behaviours**: The accelerometer time-series data was plotted alongside the annotated behaviors to visualize the alignment.
3. **Inspection of the gap based on the annotated behaviours and shape of the accelerometer time-Series**: The plot was inspected to identify any discrepancies between the annotated behaviors and the accelerometer data, with a focus on the shape and magnitude of the accelerometer time-series.
4. **Quantify the gap in seconds:** The time difference (gap) between the annotated behaviors and the accelerometer data was measured in seconds (if present).
5. **Correction of the gap**: The measured gap was carefully corrected by adjusting the accelerometer time-series data to align accurately with the annotated behaviors, ensuring the highest data accuracy.
6. **Inspect Again Before Usage**: The accelerometer time-series with adjusted timestamps was re-inspected before being included in the dataset to ensure accurate alignment with the annotated time-series.

## 7. Reliability evaluation

Two machine learning models were developed to evaluate the classification performance in two different scenarios and confirm ActBeCalf's reliability. The code is available in the folder *Machine Learning*.

### 7.1. Additional time-series classification

In addition to the original X, Y, and Z accelerometer time-series available in AcTBeCalf, five additional time-series were derived from the X, Y, and Z axes readings: Magnitude (Equation 01), ODBA (Overall Dynamic Body Acceleration), VeDBA (Vectorial Dynamic Body Acceleration), pitch, and roll. [6] provide a detailed explanation of them. (also see code *additional_ts_generation.ipynb*).

### 7.2. Split the dataset into a training and testing set

The dataset was divided into training and testing sets, with 80% of the calves (24 calves) used for model training and 20% (6 calves) used for model testing (see code *holsteinlib/genSplit.py*).

### 7.3. Modelling with machine learning models

#### 7.3.1. Model 01: Active vs Inactive classification

A total of 88 hand-crafted features (mean, median, standard deviation, min, max, first quartile, third quartile, entropy, motion variation, kurtosis, and skew) were calculated from the 8 time-series. The annotated behaviors were labeled as **active** and **inactive**. Lying and standing behaviors were categorized as inactive, while all other behaviors were labeled as active. A RandomForest model was employed for classification (see the code *active_inactive_classification.ipynb*), achieving a balanced accuracy of 0.92 (see Figure 14). The inactive class had the highest precision at 0.96, while the active class achieved the highest recall at 0.94. A detailed analysis of this work is provided by [7].

#### 7.3.2. Model 02: Four behaviour classification

This classification utilizes the mini-ROCKET feature derivation mechanism introduced by [4]. The derived ROCKET features [3] were then used to train a RidgeClassifierCV (see the code *four_behv_classification.ipynb*). The behaviors were labeled into four classes: drinking milk, lying, running, and other. The "*other*" class included all behaviors except drinking milk, lying, and running. The model performed well, achieving a balanced accuracy of 0.84 (see Figure 15). The lying behavior attained the highest precision at 0.89, while the running behavior achieved the highest recall at 0.99. A detailed analysis of this work is provided by [7].

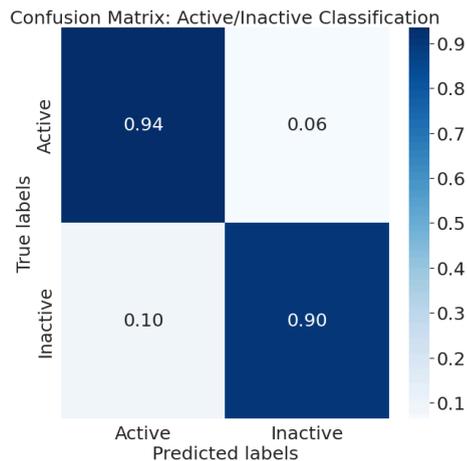
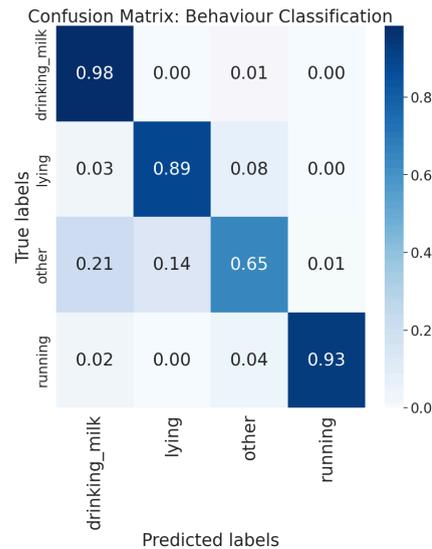

**Figure 14:** Confusion Matrix for Active vs. Inactive classification.

**Figure 15:** Confusion Matrix for 4 behaviour classification.

The sound performance obtained with these two models compared with the literature [10] attest to the reliability of AcTBeCalf for calf behavior classification from accelerometer data.

## LIMITATIONS

The primary challenge in this research is the annotation of behaviors, a process that is both time-intensive and necessitates that observers undergo joint training using an ethogram. Consequently, the volume of annotated accelerometer data is significantly smaller than the available accelerometer and video data. Several strategies could be considered to expedite the annotation process. For instance, once sufficient data has been collected to develop an initial model encompassing a few behaviors, this model could be employed to classify behaviors in unannotated accelerometer time-series. Subsequently, BORIS software could verify the consistency between these predictions and the behaviors observed in the videos.

Another main challenge is the synchronization between the accelerometer time-series and behaviors from the videos. The few lags observed when creating AcTBeCalf led us to visualize all the accelerometer time-series to make sure that the alignment was accurate, which is highly laborious and time-consuming. In addition, trials have shown that the time drift of accelerometer sensors is not systematically linear, which makes the correction very tricky. As the time drift accumulates over time, one recommendation for future studies is to equip the animals for only a few hours, then connect the accelerometers to a laptop before launching the sensors and attaching the sensors again to the animals. This reset of the real-time clock is crucial to prevent the accumulation of time drift, ensuring the accuracy of our data and the reliability of our results. Other tips would be to develop an automatic device to generate the pattern in the accelerometer signal (e.g., a mechanical arm with automatic control of shaking time, etc.). Indeed, the procedure was carried out manually by the experimenters in our study, which led to a few variabilities in the patterns generated in the accelerometer signal. The algorithm's parameters to detect the patterns automatically were, therefore, tricky to optimize, and several approaches should have been tested.

Finally, the last limitation stems from the data imbalance among behavior classes. As shown in Figure 02, lying behavior dominates the dataset. This is because pre-weaned calves housed in group pens primarily spend their time lying down. Consequently, high-movement activities like running, playing, and walking are less frequent. Additionally, less common behaviors such as urination, defecation, scratching, coughing, and jumping are observed less often. To mitigate this, active videos were identified as described in Section 5.1, allowing for a more balanced annotation of frequently displayed behaviors. However, dealing with imbalanced datasets is a common issue in applied machine learning: ActBeCalf can thus support the development of supervised classification methods for imbalanced datasets.

## ETHICS STATEMENT

The trial was conducted strictly adhering to the European Union (Protection of Animals Used for Scientific Purposes) Regulations 2012 (S.I. No. 543 of 2012). Ethical approval for the study was duly obtained from the Teagasc Animal Ethics Committee (TAEC; TAEC2021–319), ensuring all procedures involving animals were performed responsibly and ethically, in line with established guidelines.

## CRediT AUTHOR STATEMENT

- **Oshana Iddi Dissanayake**: Methodology, Software, Data Curation, Writing - Original Draft
- **Sarah E. McPherson**: Data collection
- **Joseph Allyndrée:** Data Curation
- **Emer Kennedy**: Resources, Project administration
- **Pádraig Cunningham**: Writing - Review & Editing, Supervision
- **Lucile Riaboff**: Conceptualization, Methodology, Data Curation, Writing - Review & Editing, Project administration.

## ACKNOWLEDGEMENTS

This publication has emanated from research conducted with the financial support of SFI and the Department of Agriculture, Food and Marine on behalf of the Government of Ireland to the VistaMilk SFI Research Centre under Grant Number 16/RC/3835.

## DECLARATION OF COMPETING INTERESTS

- The authors declare that they have no known competing financial interests or personal relationships that could have appeared to influence the work reported in this paper.

## REFERENCES

1. Barry, J., Kennedy, E., Sayers, R., de Boer, I.J., Bokkers, E., 2019. Development of a welfare assessment protocol for dairy calves from birth through to weaning. Animal Welfare 28, 331–344.

2. Conneely, M., Berry, D., Murphy, J., Lorenz, I., Doherty, M., Kennedy, E., 2014. Effects of milk feeding volume and frequency on body weight and health of dairy heifer calves. Livestock Science 161, 90–94.

3. Dempster, A., Petitjean, F., Webb, G.I., 2020. Rocket: exceptionally fast and accurate time series classification using random convolutional kernels. Data Mining and Knowledge Discovery 34, 1454–1495.

4. Dempster, A., Schmidt, D.F., Webb, G.I., 2021. Minirocket: A very fast (almost) deterministic transform for time series classification. In: Proceedings of the 27th ACM SIGKDD conference on knowledge discovery & data mining, pp 248–257.

5. Dissanayake, O., McPherson, S., Allyndree, J., Kennedy, E., Cunningham, P., Riaboff, L., 2023. Personalized Weighted AdaBoost for Animal Behavior Recognition from Sensor Data. In: 2023 31st Irish Conference on Artificial Intelligence and Cognitive Science (AICS). IEEE, pp. 1-8.

6. Dissanayake, O., McPherson, S.E., Allyndree, J., Kennedy, E., Cunningham, P., Riaboff, L., 2024. Evaluating ROCKET and Catch22 features for calf behaviour classification from accelerometer data using Machine Learning models. arXiv. https://doi.org/10.48550/arXiv.2404.18159.

7. Dissanayake, O., McPherson, S.E., Allyndrée, J., Kennedy, E., Cunningham, P., Riaboff, L., 2024. Development of a digital tool for monitoring the behaviour of pre-weaned calves using accelerometer neck-collars. arXiv. https://doi.org/10.48550/arXiv.2406.17352.

8. Friard, O., Gamba, M., 2016. Boris: a free, versatile open-source event logging software for video/audio coding and live observations. Methods in Ecology and Evolution 7, 1325–1330.

9. Lubba, C.H., Sethi, S.S., Knaute, P., Schultz, S.R., Fulcher, B.D., Jones, N.S., 2019. catch22: Canonical time-series characteristics: Selected through highly comparative time-series analysis. Data Mining and Knowledge Discovery 33, 1821–1852.

10. Riaboff, L., Shalloo, L., Smeaton, A.F., Couvreur, S., Madouasse, A., Keane, M.T., 2022. Predicting livestock behaviour using accelerometers: A systematic review of processing techniques for ruminant behaviour prediction from raw accelerometer data. Computers and Electronics in Agriculture 192, 106610.

11. Rushen, J., Chapinal, N., de Passile, A., 2012. Automated monitoring of behavioural-based animal welfare indicators. Animal Welfare 21, 339–350.

12. van Hees, V.T., Gorzelniak, L., Dean Leon, E.C., et al., 2013. Separating movement and gravity components in an acceleration signal and implications for the assessment of human daily physical activity. PLoS One 8(4), e61691. https://doi.org/10.1371/journal.pone.0061691.

# APPENDEX

**Table A1:** Each behavior, its associated modifier, the total duration, the number of segments (observations), and the total number of calves the behavior was observed from.

| Behavior | Modifier | duration(seconds) | number of segments | number of calves |
|---|---|---|---|---|
| Standing | | 7532.42 | 737 | 30 |
| Lying | | 39021.56 | 120 | 27 |
| Drinking | electrolytes | 32.91 | 3 | 3 |
| | milk | 8620.60 | 169 | 27 |
| | water | 75.76 | 5 | 2 |
| Eating | bedding | 5139.85 | 82 | 18 |
| | concentrates | 5871.62 | 84 | 21 |
| | forage | 3202.15 | 44 | 17 |
| Walking | forward | 2657.58 | 561 | 30 |
| | backward | 296.17 | 65 | 20 |
| Run | | 3306.04 | 608 | 25 |
| Grooming | standing | 4515.58 | 334 | 29 |
| | lying | 476.67 | 22 | 6 |
| | groom \| standing | 1187.58 | 38 | 15 |
| | groom \| lying | 113.32 | 4 | 3 |
| Social interaction | nudge \| standing | 223.89 | 29 | 13 |
| | nudge \| lying | 6.99 | 1 | 1 |
| | sniff \| standing | 914.57 | 114 | 25 |
| | sniff \| lying | 97.38 | 5 | 3 |
| Play | headbutt | 701.66 | 65 | 13 |
| | object | 214.12 | 10 | 6 |
| | jump | 72.45 | 24 | 8 |
| | mount | 7.76 | 3 | 2 |
| Rising | | 258.59 | 55 | 25 |
| Lying down | | 193.42 | 43 | 22 |
| Rumination | standing | 24.77 | 2 | 2 |
| | lying | 660.59 | 21 | 5 |
| Defecation | | 76.83 | 6 | 5 |
| Urination | | 230.50 | 10 | 9 |
| Oral manipulation of pen | | 4713.86 | 158 | 24 |
| Sniff | standing | 5362.86 | 380 | 29 |
| | lying | 197.68 | 7 | 5 |
| | walking | 454.78 | 52 | 20 |
| Abnormal | cross-suckle_udder | 67.70 | 2 | 1 |
| | cross-suckle_other | 34.97 | 1 | 1 |
| | tongue rolling \| standing | 234.87 | 24 | 9 |
| | tongue rolling \| lying | 11.97 | 1 | 1 |
| SRS | scratch | 404.63 | 60 | 20 |
| | rub | 265.18 | 20 | 11 |
| | stretch | 55.74 | 9 | 9 |
| Cough | | 17.15 | 4 | 3 |
| Fall | | 19.32 | 6 | 5 |
| Vocalisation | | 3.23 | 1 | 1 |